\documentclass[twocolumn,showpacs,showkeys,preprintnumbers,amsmath,amssymb]{revtex4}

\usepackage{graphicx}

\begin{document}

\title{Charge redistribution at YBa$_2$Cu$_3$O$_7$-metal interfaces}

\author{U.~Schwingenschl\"ogl}
\author{C.~Schuster}
\affiliation{Institut f\"ur Physik, Universit\"at Augsburg, 86135 Augsburg, Germany}

\date{\today}

\begin{abstract}
Charge redistribution at interfaces is crucial for electronic
applications of high-T$_c$ superconductors, since the band structure
is modified on a local scale. We address the normal-state electronic
structure of YBa$_2$Cu$_3$O$_7$ (YBCO) at an YBCO-metal contact by
first principles calculations for prototypical
interface configurations. We derive quantitative
results for the intrinsic doping of the superconducting CuO$_2$
planes due to the metal contact. Our findings can be explained in
terms of a band-bending mechanism, complemented by local screening
effects. We determine a net charge transfer of 0.09 to 0.13 electrons
in favour of the intraplane Cu sites, depending on the interface
orientation.
\end{abstract}



\maketitle

Electronic transport in wires and tapes from high-T$_c$ materials
is seriously affected by structural defects and interfaces. In
contrast to most conventional superconductors, bending of the band
structure due to local variations of the charge distribution is
strong enough to control the transport properties
\cite{mannhart98,hilgenkamp02}. This results from large dielectric
constants and small carrier densities, characteristic for high-T$_c$
materials \cite{samara90}. The Thomas-Fermi screening length, over
which band-bending is effective, therefore reaches the order of
magnitude of the superconducting coherence length.

The specific contact resistivity of YBCO-metal thin films is known
to depend on the details of the contact geometry \cite{hahn94}. In
particular, the transport in micron-sized YBCO-metal heterojunctions
is strongly affected by the orientation of the YBCO crystallographic
axes with respect to the direction of the current flow
\cite{komissinskii01}. From the theoretical point of view, effects
of charge modulation at the surface of high-T$_c$ superconductors
have been studied by Emig {\it et al.} \cite{emig97}. It turns out,
that surfaces are covered by dipole layers, due to a local
suppression of the gap function. Nikolic {\it et al.} \cite{nikolic02}
study the charge imbalance at the boundary between a short coherence
length superconductor and a normal metal by means of a self-consistent
microscopic approach. However, first principles electronic structure
calculations taking into account the details of the crystal structure
are missing so far, probably due to a high demand on CPU time.

The technical optimization of interfaces calls for insight into the
electronic structure close to the contact, which we address in the
following for characteristical YBCO-metal interfaces. Since the
electronic properties depend on the local atomic configuration, it is
necessary to start from the details of the crystal structure in order
to obtain reliable results. We present findings of band structure
calculations for two prototypical YBCO-metal interface configurations
fulfilling this requirement. Our calculations are based on density
functional theory and the generalized gradient approximation, as
implemented in the WIEN$2k$ program package \cite{wien2k}. This
full-potential linearized augmented-plane-wave code is known to be
particularly suitable for dealing with structural relaxation and
charge redistribution in complex geometrical arrangements
\cite{us07a, us07b, us07c}. We obtain quantitative results for the
band-bending magnitude and, therefore, for the intrinsic doping of
the superconducting CuO$_2$ planes due to the metal contact.

\begin{figure}
\includegraphics[width=0.45\textwidth,clip]{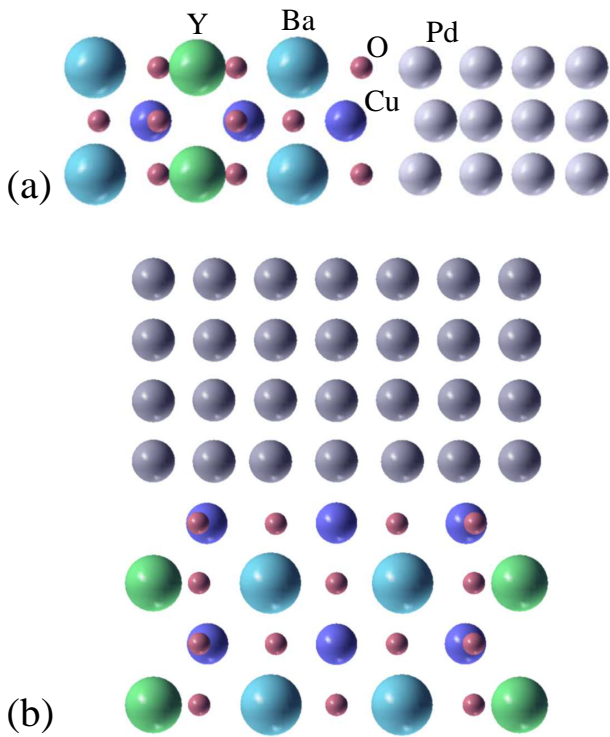}
\caption{YBCO-metal interface configurations: (a) parallel and (b)
perpendicular. The naming refers to the orientation of the interface with
respect to the CuO$_2$ planes.}
\label{fig1}
\end{figure}

Since band-bending is proposed to take place on the length scale of
the YBCO lattice constant, the electronic structure of YBCO-metal
interfaces becomes accessible to a supercell approach with periodic
boundary conditions. In the following, we address supercells both
parallel and perpendicular to the crystallographical $c$-axis. In
each case, we start from the experimental YBCO bulk lattice constants 
\cite{siegrist87} and optimize the atomic coordinates in order to
minimize the atomic forces. In a second step, the structural relaxation
of the supercells is carried out \cite{kouba97,kouba99}. Figure
\ref{fig1} shows the interface configurations under consideration,
which we call the parallel (a) and the perpendicular (b) interface,
referring to their orientation with respect to the CuO$_2$ planes.

It is convenient to choose fcc Pd as the metallic substituent,
because of a minimal lattice mismatch of about 0.7\%. The supercell
for the parallel interface consists of 2 YBCO unit cells, whereas 3
YBCO unit cells are used for the perpendicular interface. The YBCO
domain in each case terminates with a Cu-O layer \cite{xin89,edwards92,derro02}.
Furthermore, the metal domain comprises 3 and 4 Pd fcc unit cells
for the perpendicular and parallel configuration, respectively.
Relative shifts between the YBCO and metal domain parallel to the
interface have not be taken into consideration, since they do not
affect our further conclusions \cite{us07b}. Structure optimization
results in a strong tendency towards Pd-O bonding, whereas repulsion
is found between Cu and Pd atoms. Importantly, the structural
relaxation affects almost only the first atomic layer off the contact.
Bond lengths calculated for these sites are summarized in Table
\ref{tab1} for both supercells.

\begin{table}
\caption{Selected bond lengths at the parallel and perpendicular
YBCO-metal interface, as resulting from the structure optimization.}
\begin{tabular}{l|c|c}
&parallel & perpendicular\\\hline
d$_{\rm Cu-Pd}$&3.32\,\AA, 3.79\,\AA&2.62\,\AA, 2.74\,\AA, 2.77\,\AA\\
d$_{\rm O-Pd}$&2.12\,\AA, 3.87\,\AA&2.02\,\AA, 3.15\,\AA\\
d$_{\rm Cu-O}$&2.02\,\AA&1.94\,\AA\\
d$_{\rm Cu-O_{Ba}}$&1.90\,\AA&1.93\,\AA
\end{tabular}
\vspace{0.5cm}
\label{tab1}
\end{table}

\begin{figure}
\includegraphics[width=0.45\textwidth]{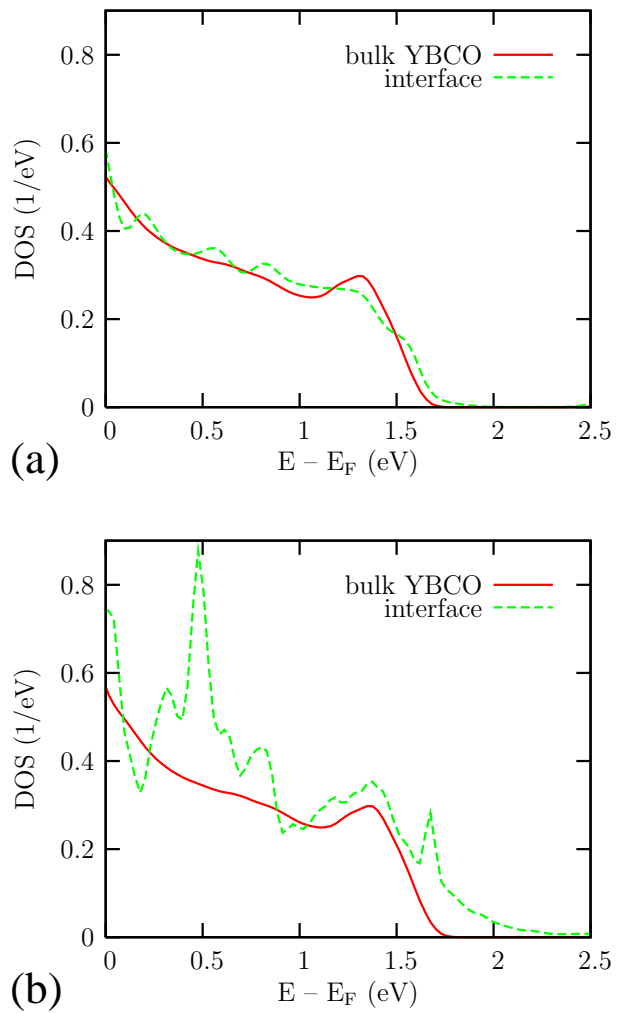}
\caption{Partial Cu $3d$ densities of states for CuO$_2$-plane sites
in the vicinity of the parallel (a) and perpendicular (b) interface.
The gross shape of the interface DOS resembles the bulk DOS when the
latter is shifted by 0.20\,eV (a) and 0.15\,eV (b) to lower energies,
respectively, which corresponds to charge carrier reductions of 0.13
and 0.09 holes per Cu site.}
\label{fig3}
\end{figure}

In the following discussion, we compare the electronic structures obtained for
our YBCO-metal interfaces with the results of a bulk YBCO calculation.
We therefore mention that our bulk YBCO density of states (DOS) agrees
perfectly with previous theoretical and experimental findings,
see \cite{pickett89,pickett90,wechsler97} and the references given
therein. Figure \ref{fig3} shows partial Cu $3d$ DOS curves for Cu
sites in the CuO$_2$-planes of bulk YBCO and our interfaces (a) and
(b). In the latter cases, the results refer to the Cu sites second closest to
the contact. As compared to bulk YBCO, additional Cu states appear
in the vicinity of the Fermi energy for the Cu atoms actually
forming the interface, which trace back to a modified Cu-O bonding.
In contrast, the second closest Cu atoms resemble the bulk YBCO DOS
very well and hence allow us to study the band-bending. For the
parallel interface, we observe almost perfect agreement of the DOS
curves, whereas some effects of the structural relaxation are left
for the perpendicular interface. However, in both cases the bulk DOS
has to be shifted to lower energies in order to reconcile the curves.
The necessary shift amounts to 0.20\,eV for the parallel interface,
whereas 0.15\,eV are sufficent in the perpendicular case. 

Since the gross structure of the Cu $3d$ DOS is not affected by the
interface, we can interpret these energetical shifts in terms
of almost ideal down bending of the electronic bands due to a
modified Fermi level. As a consequence, the hole count at the Cu 
sites is altered. To be specific, a shift of 0.20\,eV corresponds to
a reduction of 0.13 holes, and a shift of 0.15\,eV comes along with a
loss of 0.09 holes. Both these values and the calculated magnitudes
of the band-bending are expected to be independent of the metallic
substituent used for the interface, which we have confirmed for
silver contacts. Importantly, the charge transfer likewise depends
only little on the orientation of the YBCO-metal interface with
respect to the unit cell of the high-T$_c$ compound. An intrinsic
doping close to 0.1\,eV hence appears to be a general result for
YBCO-metal contacts. Of course, oxygen defects close to the interface
could modify the charge transfer.

Core levels at atomic sites near the interface show energetical
shifts of about twice the magnitude reported for the Cu valence
states. To understand this fact, electronic screening has
to be taken into consideration. Screening is more efficient for the
Cu $3d$ states than for any core states, since the former have
finite weight at the Fermi energy. In general, the electrostatic
screening length is only a few nanometers in high-T$_c$ cuprates and
the crystal structure is very inhomogeneous. Conventional band-bending
models based on a continuum description of the charge distribution
thus cannot be applied. Nevertheless, electronic screening results
in a significant reduction of the band-bending magnitude at YBCO-metal
interfaces, therefore in a reduced charge transfer. 

Xu and Ekin report on specific resistivities for YBCO-Au interfaces
of $10^{-4}\,\Omega\,{\rm cm}^2$ to $10^{-3}\,\Omega\,{\rm cm}^2$ at
low temperatures \cite{xu04}. However, our calculations for normal-state
YBCO-metal interfaces do not show a relevant reduction of the Cu $3d$ DOS
at the Fermi energy. Even though no insulating layer is formed in the
vicinity of the YBCO-metal contact, the observed interface resistivity
can be explained in terms of screened band-bending. Since the charge
carrier density in the CuO$_2$-planes is significantly reduced, a
local breakdown of the superconductivity is to be expected.

We have presented electronic structure calculations for prototypical
contact configurations between the short coherence length
superconductor YBa$_2$Cu$_3$O$_7$ and a normal metal. In particular,
we have discussed the charge redistribution in high-T$_c$ materials
induced by normal metal interfaces. Calculations for well-relaxed
supercells show that the charge redistribution can be interpreted in
terms of an intrinsic doping of the superconductor on a nanometer
length scale, in correspondence with the experimental observation of
charge carrier depletion. The net charge transfer in favour of the
copper sites in the CuO$_2$-planes amounts to 0.13 electrons when
the interface is oriented parallel to the superconducting planes, and
to 0.09 electrons for the perpendicular orientation.

This weak dependence of the charge transfer magnitude on the
orientation of the interface lets expect that the intrinsic doping at
YBCO-metal interfaces is given by some 0.1 electrons for any contact
geometry, as long as the YBCO domain terminates with a Cu-O layer.
Moreover, the mechanism of screened band-bending and the net charge
transfer are almost independent of the specific high-T$_c$ material
and metal forming the contact. Therefore, the results are very
general and can be applied to a large variety of interfaces.

\begin{acknowledgments}
We acknowledge valuable discussions with U.\ Eckern, V.\ Eyert, J.\ Mannhart,
and T.\ Kopp, and financial support by the Deutsche Forschungsgemeinschaft
(SFB 484).
\end{acknowledgments}

\end{document}